\documentclass[%
reprint,
amsmath,amssymb,
aps,
pra,
]{revtex4-1}

\usepackage{graphicx}
\usepackage{hyperref}
\usepackage{dcolumn}
\usepackage{bm}
\usepackage{color}
\usepackage{soul}

\usepackage{enumitem}

\newcommand{\ket}[1]{\left| #1 \right>} 
\newcommand{\braket}[2]{\left< #1 \vphantom{#2} \right| \left. #2 \vphantom{#1} \right>} 

\begin{document}

\preprint{APS/123-QED}

\title{Coupled modes locally interacting with qubits: critical assessment of the rotating wave approximation}
\author{P. C. C\'ardenas}
\affiliation{Departamento de F\'isica y Matem\'aticas, Universidad Aut\'onoma de Manizales, Antigua Estaci\'on del Ferrocarril, Manizales, Caldas, Colombia}
\author{W. S. Teixeira}
\affiliation{Centro de Ci\^encias Naturais e Humanas, Universidade Federal do ABC, Santo Andr\'e, 09210-170 S\~ao Paulo, Brazil}
\author{F. L. Semi\~ao}
\affiliation{Centro de Ci\^encias Naturais e Humanas, Universidade Federal do ABC, Santo Andr\'e, 09210-170 S\~ao Paulo, Brazil}

%


\date{\today}

\begin{abstract}
The interaction of qubits with quantized modes of electromagnetic fields has been largely addressed in the quantum optics literature under the rotating wave approximation (RWA), where rapid oscillating terms in the qubit-mode interaction picture Hamiltonian can be neglected. At the same time, it is generally accepted that provided the interaction is sufficiently strong or for long times, the RWA tends to describe physical phenomena incorrectly. In this work, we extend the investigation of the validity of the RWA to a more involved setup where two qubit-mode subsystems are brought to interaction through their harmonic coordinates. Our treatment is all analytic thanks to a sequence of carefully chosen unitary transformations which allows us to diagonlize the Hamiltonian within and without the RWA. By also considering qubit dephasing, we find that the purity of the two-qubit state presents non-Markovian features which become more pronounced as the coupling between the modes gets stronger and the RWA loses its validity. In the same regime, there occurs fast generation of entanglement between the qubits which is also not correctly described under the RWA. The setup and results presented here clearly show the limitations of the RWA in a scenario amenable to exact description and free from numerical uncertainties. Consequently, it may be of interest for the community working with cavity or circuit quantum electrodynamic systems in the strong coupling regime.
\end{abstract}

\maketitle


\section{\label{sec:level1}Introduction}
Modeling physical phenomena via the coupling of two-level systems (qubits) to quantized harmonic oscillators has historically been of great interest in diverse fields ranging from quantum optics \cite{kni05,jc63,ebly80,kni82} and solid-state physics \cite{mah90, wei08} to quantum biology \cite{sch15,qbio13,ore14,gil05}. This approach has become frequent in modern physics since fully quantum-mechanical descriptions may reveal phenomena not covered by classical or semiclassical approaches. As well-known examples, one has the study of dissipation and decoherence of a qubit via the spin-boson model \cite{leg87} or the presence of collapses and revivals of the atomic population inversion in the atom-field interaction \cite{ebly80, kni82}.

For the latter case, the simplest quantum description is given by the exactly solvable Jaynes-Cummings (JC) model \cite{jc63}, which considers a two-level atom weakly coupled to a single mode of the electromagnetic field. Usually, the ``energy non-conserving' terms in the atom-field interaction Hamiltonian are neglected through the so-called rotating wave approximation (RWA). However, the use of the RWA in this problem may not describe dynamical properties of the model correctly when the atom-field interaction becomes sufficiently strong \cite{zub87, nad11}. Other studies have addressed the limitations of the RWA in diverse configurations \cite{hu10,hau10,hau08,aga71,liao16,sor04,ebly12}.  Only recently it has been shown that the non-RWA atom-field Hamiltonian possesses a symmetry rendering the model integrable \cite{Braak}. However, due to the lack of closed form expressions for the eigenstates, one usually has to appeal to different effective approaches to treat the problem without the RWA. This includes the use perturbation series for path-integrals \cite{zub87} or particular regimes such as far-from-resonance cases (dispersive limit) \cite{zue09}. 

Recently, the fabrication of artificial atoms with superconducting circuits \cite{cla08,gir08,wall04,bla04,bla07,scho07} has favored the control of qubit-oscillator interactions, and thus regimes where the RWA breaks down can be explored experimentally. In circuit quantum electrodynamics (circuit QED) a qubit or two-level system can be produced, for instance, by using a thin insulator between two superconducting materials [Josephson junction (JJ)], and controlling either the number of Cooper pairs that tunnel from one side to the other (charge qubit) or the phase of their wavefunctions (phase qubit). Also, by adding one or more JJ in a superconducting loop [Superconducting Quantum Interference Device (SQUID)], a qubit can be produced by controlling the external magnetic flux through the SQUID (flux qubit) \cite{cla08,gir08}. On the other hand, a quantum harmonic oscillator naturally represents a single electromagnetic mode trapped in a transmission line \cite{gir08,bla04}.

Giving all these developments in the control of simple systems consisting of qubits and bosonic modes, it is natural to search for configurations which allow us to further understand the limits of the usually taken RWA.  We explore the validity of the RWA in a setup comprised of two identical qubit-oscillator systems that are coupled through their harmonic coordinates. This setup is amenable to implementation in superconducting circuits as discussed in  \cite{car15}, where non-Markovian features are discussed within the RWA. The present work is organized as follows. We analytically diagonalize the full non-RWA Hamiltonian in (Sec.~\ref{subsec:diag}) and then solve the master equation which includes dephasing for the qubits (Sec.~\ref{subsec:dyn}). Predictions contrasting our full treatment with the RWA model in  \cite{car15} are then presented, in particular for the two-qubit subsystem for which purity (Sec.~\ref{subsec:pur}) and entanglement dynamics (Sec.~\ref{subsec:ent}) are investigated. Finally, in Sec.~\ref{sec:conc} we present our conclusions.
\section{\label{sec:model}The model}
\begin{figure}[b!]
\includegraphics[scale=0.4]{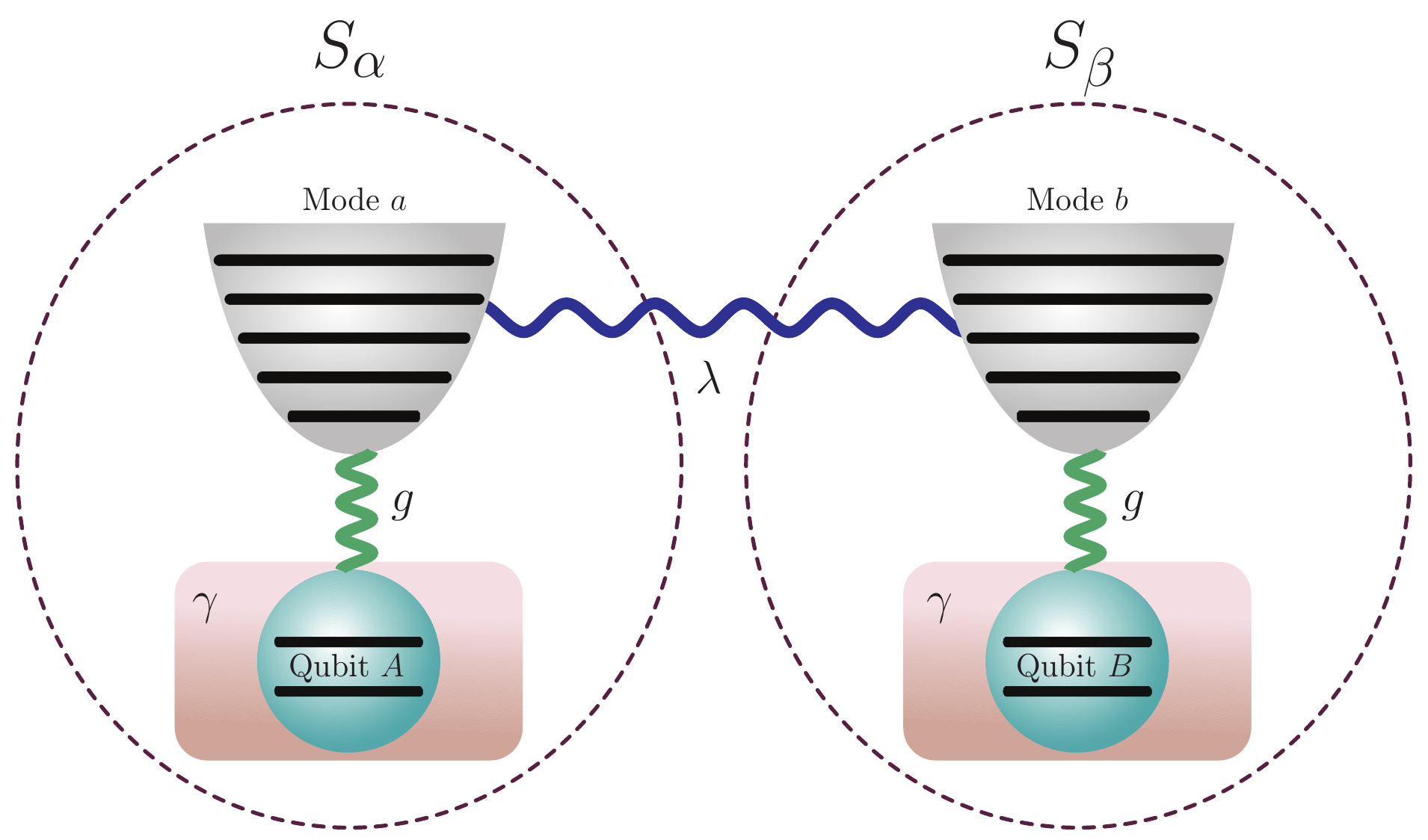}
\caption{\label{fig1} (Color online) Two subsystems $S_\alpha$ and $S_\beta$, each formed by a qubit and a bosonic mode, interact through the modes (coupling strength $\lambda$). Each qubit is also subjected to dephasing at rate $\gamma$. Inside each subsystem, the qubit and the mode interact with each other (coupling strength $g$). }
\end{figure}

In this work, we are interested in the setup depicted in Fig.\ref{fig1}. It consists of two groups of subsystems ($\alpha$ and $\beta$), each comprising a qubit and a bosonic mode. These groups will interact through the modes. Inside each group, the local interaction takes the usual spin-boson form with no transverse field ($\hbar=1$)
\begin{equation}
H_{\text{SB}} = g\left(\sigma_{z_A}+\mu I_A\right)\left(a^{\dagger}+a\right)+g\left(\sigma_{z_B}+\mu I_B\right)\left(b^{\dagger}+b\right),
\label{eq:sbH}
\end{equation}
where $\sigma_{z_{A(B)}}$ are the Pauli matrices for the qubits; $a^{\dagger}(a)$ and $b^{\dagger}(b)$ are the creation (annihilation) operators for the corresponding bosonic modes, $g$ is a coupling constant, and $I_{A(B)}$ is the identity operator acting on the corresponding qubit state space. The $\mu$-terms  naturally appear in some circuit QED architectures when working out of the so-called degeneracy point \cite{bla04,bla07}, so that, for completeness, they are included here. More details can be found in \cite{car15}.

The total energy of the full setup reads
\begin{equation}
H_j = H_0+H_{\text{SB}}+H_{\text{BB}_j},
\label{eq:genH}
\end{equation}
with the free Hamiltonian 
\begin{equation}
H_0 = \frac{\omega_0}{2}\left(\sigma_{z_A}+\sigma_{z_B}\right)+\omega \left(a^{\dagger}a+b^{\dagger}b\right),
\label{eq:freeH}
\end{equation}
where $\omega_0$ is the resonance frequency of the qubits and $\omega$ is the angular frequency of the modes. The index $j\in\{1,2\}$ defines the form of the interaction mechanism between the modes which can be either
\begin{equation}
H_{\text{BB}_1} = \lambda\left(a^{\dagger}+a\right)\left(b^{\dagger}+b\right)
\label{eq:bb1H}
\end{equation}
or
\begin{equation}
H_{\text{BB}_2} = \lambda\left(a^{\dagger}b+a b^{\dagger}\right),
\label{eq:bb2H}
\end{equation}
with $\lambda$ being a coupling constant. In the first approach ($j=1$), the spatial coordinates of each oscillator are coupled in a kind of quadrature-quadrature form. In circuit QED,  this can be induced by coupling the two transmission lines to an auxiliary qubit that mediates a geometric second-order interaction \cite{sol08}, while in cavity QED this can be done by placing a partially reflecting mirror between two optical cavities \cite{har06}. On the other hand, the second approach ($j=2$) arises from the RWA performed on $H_{\text{BB}_1}$ so that its oscillating terms are neglected in the interaction picture. For the interaction of a two-level atom with an electromagnetic mode, which in the RWA gives rise to the JC Hamiltonian, one can only compare the RWA and non-RWA Hamiltonians through successive approximations or numerics. Here, we will be able to perform such investigation in a fully analytic manner by exactly solving
\begin{equation}
\dot{\rho_j}=-i\left[H_j,\rho_j\right]+\frac{\gamma}{2}\left(\sigma_{z_A}\rho_j\sigma_{z_A}+\sigma_{z_B}\rho_j\sigma_{z_B}-2\rho_j\right),
\label{eq:msteq}
\end{equation}
for initial states of interest. In Eq.(\ref{eq:msteq}),  $\gamma$ is the rate of pure dephasing caused by independent Markovian baths acting on the qubits. This is by far the most relevant noise when working outside the degeneracy point \cite{dp1,dp2}. Energy relaxation of qubits or the transmission lines (bosonic modes), as well as dephasing on the latter, can be made negligible compared to dephasing in the qubits \cite{bla04,bla07}. These experimental facts provide us a backgound to leave aside noise mechanisms besides qubit dephasing as a first approximation. This is especially convenient in our case because our goal is to provide analytical expressions that evidence inadequacy of the RWA in certain regimes. Those neglected noise mechanisms would render the problem unsuitable to analytic treatment and can be numerically investigated elsewhere.
\section{\label{sec:res}Results}
\subsection{\label{subsec:diag}Diagonalization}
In order to  analytically solve Eq.\eqref{eq:msteq}, we start by diagonalizing the Hamiltonians $H_j$. This is achieved by the unitary transformation $U_j$ given by $U_j=P_jS_jTD_j$ where
\begin{eqnarray}
D_j=e^{\delta_j\left(a^{\dagger}-a+b^{\dagger}-b\right)}
\end{eqnarray}
is the displacement operator with $\delta_j=g\mu/\left(\omega+2^{2-j}\lambda\right)$,
\begin{equation}
T=e^{\frac{\pi}{4}\left(a^{\dagger}b-ab^{\dagger}\right)}
\label{eq:t2}
\end{equation}
corresponds to a beam-splitter operation,
\begin{equation}S_j=e^{-\frac{r_{j_+}}{2}\left(a^2-a^{\dagger 2}\right)}e^{-\frac{r_{j_-}}{2}\left(b^2-b^{\dagger 2}\right)}
\label{eq:t3}
\end{equation}
is a squeezing operator with $r_{1_{\pm}}=\ln\left(1\pm 2\lambda/\omega\right)$ and $r_{2\pm}=0$, and
\begin{equation}
P_j=e^{\lambda_{j_+}\left(\sigma_{z_B}+\sigma_{z_A}\right)\left(a^{\dagger}-a\right)}e^{\lambda_{j_-}\left(\sigma_{z_B}-\sigma_{z_A}\right)\left(b^{\dagger}-b\right)}
\label{eq:t4}
\end{equation}
is a polaron transformation \cite{pol} and $\lambda_{j_\pm}=g e^{-r_{j_\pm}}/(\sqrt{2}\Omega_{j_\pm})$. Notice that the operation $S_j$ reduces to the identity for $j=2$. This is so because this transformation is responsible for the elimination of terms with $ab$ and $a^{\dagger}b^{\dagger}$, not present in Eq.~\eqref{eq:bb2H}. It is important to realize that the application of $S_1$ requires $\lambda<\omega/2$, so that the frequencies of the normal modes are real numbers. Values of $\lambda$ close to such limit have been associated to quantum chaos in nonlinear oscillators \cite{zue14}. 

\begin{figure}[h!]
\includegraphics[scale=0.52]{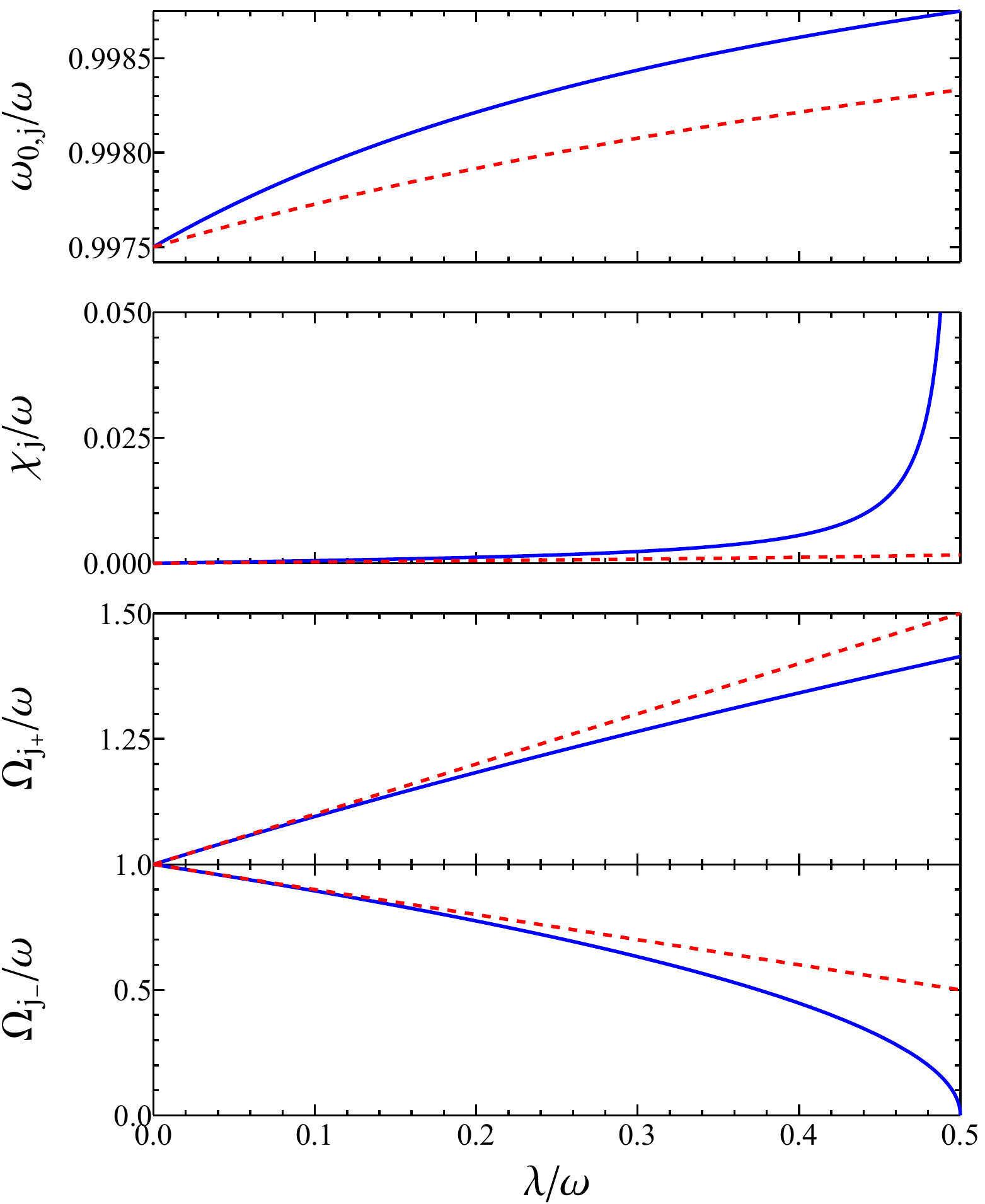}
\caption{\label{fg:Hjcomp} (Color online) Parameters of the diagonal Hamiltonian~\eqref{eq:diagHj} as functions of $\lambda$ (in units of $\omega$). \textbf{First row}: Effective frequencies of the qubits. \textbf{Second row}: Effective coupling strength of the qubits. \textbf{Third and forth rows}: Effective frequencies of the modes. In all cases, the solid blue lines represent the complete model ($j=1$), whereas the dashed red lines represent the RWA model ($j=2$). We used $\omega_0=\omega$, $g=0.025\ \omega_0$, and $\mu=1$.}
\end{figure}

With the help of these transformations, we obtain the diagonal Hamiltonian $H_j'=U_jH_jU_j^\dag$ that reads
\begin{eqnarray}
H_j'=\frac{\omega_{0,j}}{2}\left(\sigma_{z_A}+\sigma_{z_B}\right)+\frac{\chi_j}{2}\sigma_{z_A}\sigma_{z_B}
+ \Omega_{j_+} a^{\dagger}a+\Omega_{j_-} b^{\dagger}b,\nonumber\\
\label{eq:diagHj}
\end{eqnarray}
with shifted qubit frequencies 
\begin{eqnarray}
\omega_{0,j}=\omega_0-4g\delta_j
\label{eq:shiftw0}
\end{eqnarray}
and normal mode frequencies
\begin{eqnarray}
\Omega_{j_\pm}=\omega \cosh\left(2r_{j_\pm}\right)\pm\lambda e^{-2r_{j_\pm}}.
\label{eq:shiftw}
\end{eqnarray} 
The Hamiltonian  (\ref{eq:diagHj})  is an interesting physical result. It implies that, in spite of the model ($j=1$ or $j=2$), the modes decouple from the qubits, and the latter  interact through an Ising-type Hamiltonian with
\begin{eqnarray}
\chi_j=2g^2\left(\frac{e^{-2r_{j_-}}}{\Omega_{j_-}}-\frac{e^{-2r_{j_+}}}{\Omega_{j_+}}\right).
\label{eq:chi}
\end{eqnarray}
For uncoupled modes ($\lambda=0$), no effective coupling between the qubits is observed ($\chi_j=0$). For finite  $\lambda$, we can already spot the fundamental differences in the non-RWA ($j=1$) and RWA ($j=2$) descriptions.  This can be seen from the plots in Fig.~\ref{fg:Hjcomp}, where we present the dependence of $\omega_{0,j}$, $\chi_j$, and $\Omega_{j_\pm}$ on the modes coupling constant $\lambda$. These physical frequencies clearly indicate that the RWA dismally fails with the increase of $\lambda$. 
\subsection{\label{subsec:dyn}Dynamics} 
In the space of Hamiltonian (\ref{eq:diagHj}), the system density matrix is given by $\rho_j'=U_j\rho_jU_j^\dag$. A new transformation defined as $\rho_{I_j}=e^{iH_j't}\rho_j'e^{-iH_j't}$ finally allows one to rewrite the master equation \eqref{eq:msteq} in a very compact form as
\begin{equation}
\dot{\rho_{I_j}}=\frac{\gamma}{2}\left(\sigma_{z_A}\rho_{I_j}\sigma_{z_A}+\sigma_{z_B}\rho_{I_j}\sigma_{z_B}-2\rho_{I_j}\right).
\label{eq:msteqIP}
\end{equation}
Although the modes do not appear explicitly in Eq.~\eqref{eq:msteqIP}, they have not yet been traced out. What happens is that they are \textit{frozen} in this interaction picture and were completely decoupled from the qubits due to the transformation $U_j=P_jS_jTD_j$. Consequently, we can solve Eq.~\eqref{eq:msteqIP} in the qubits subspace and tensor the result with the initial state of the modes. The transformations back to the original picture will then restore the time evolution of the whole system, entangling modes and qubits. By denoting $\ket{\psi}_A\otimes\ket{\phi}_B$ as just $\ket{\psi\phi}$, we then solve Eq.~\eqref{eq:msteqIP} in the standard basis $\{\ket{ee},\ket{eg},\ket{ge},\ket{gg}\}$, in which $\ket{e(g)}$ stands for the excited (ground) state of a single qubit. 
In our case, we do not have to move the system state entirely back to the original picture because we are interested in the dynamics of the qubits. The  polaron operation $P_j^{\dagger}$ is the only one required as the other unitary transformations employed to diagonalize Eq.~\eqref{eq:genH} are in fact local in the modes. The density matrix of the qubits is then obtained as
\begin{eqnarray}
\rho_{AB_j}(t)=\text{Tr}_{ab}\left[P_j^{\dagger}e^{-iH_j't}\rho_{I_j}e^{iH_j't}P_j\right],
\label{eq:rhoAB}
\end{eqnarray}
where the partial trace is taken over the modes degrees of freedom. 

From now on, we will focus on a particular choice of initial states that suitably illustrates the dynamics of state purity and entanglement for the qubits.  On their own, these two quantities  carry a lot of information and their analysis is then of general importance in quantum information. More important to us, they both depend on the whole density matrix and not only on its diagonal elements. Quantities such as occupation probabilities of the bare states would depend only on diagonal density matrix elements. For all these reasons, entropy and entanglement are then very good candidates to spot the differences between the full model and its RWA version. We consider the qubits to be initially prepared in the eigenstate of the Pauli matrix $\sigma_{x_{A(B)}}$ associated with the eigenvalue $+1$, i.e., 
\begin{eqnarray}
|+\rangle_A\otimes|+\rangle_B\equiv\ket{++}=\frac{1}{2}\left(\ket{ee}+\ket{eg}+\ket{ge}+\ket{gg}\right).\nonumber\\
\label{eq:i2qb}
\end{eqnarray}
On the other hand, the modes are initially set in the product of coherent states, 
\begin{eqnarray}
\ket{\alpha}_a\otimes\ket{\beta}_b\equiv\ket{\alpha}\ket{\beta}=e^{-\frac{|\alpha|^2+|\beta|^2}{2}}\sum_{n,m=0}^{\infty}{\frac{\alpha^{n}\beta^{m}}{\sqrt{n!m!}}\ket{nm}},\nonumber\\
\label{eq:i2md}
\end{eqnarray}
where $\alpha$ and $\beta$ are complex amplitudes, and $\ket{n}\otimes\ket{m}\equiv\ket{nm}$ is the two-mode Fock state. Solving Eq.~\eqref{eq:msteqIP} for the initial state $\ket{\psi(0)}=\ket{++}\ket{\alpha}\ket{\beta}$ and using Eq.\eqref{eq:rhoAB}, we obtained the $16$ components of $\rho_{AB_j}(t)$ in the standard basis as
\begin{eqnarray}
\rho_{AB,mn_j}(t)=\frac{1}{4}\gamma_{mn_j}(t)\Theta_{mn,j_+}(t)\Theta_{mn,j_-}(t)
\label{eq:kmn}
\end{eqnarray}
 with $m,n\in\{ee,eg,ge,gg\}$ and $m\ne n$. The diagonal elements $m=n$ are time-independent and given by $\rho_{AB,mm_j}=1/4$. The factors $\gamma_{mn_j}(t)$ arise from the non-unitary dynamics followed by each qubit (a dephasing factor), whereas $\Theta_{mn,j_\pm}(t)$ are scalar products originated from the partial trace operations. Explicit expressions for $\gamma_{mn_j}(t)$ and details about $\Theta_{mn,j_\pm}(t)$ can be found in the Appendix.
\subsection{\label{subsec:pur}Purity of the two-qubit subsystem}
In general, quantum information processing requires that pure states, like superposition states, remain pure during time evolution. However, when a quantum system is interacting with others, its reduced dynamics will, in general, affect the state purity. The same is valid when the system is in contact with a bath and, in this case, the lost of purity is typically irreversible.  A good measure of how much a state $\rho$ is pure in a $d$-dimensional  state space is given by a quantity called purity, defined as $\mathcal{P}=\text{Tr}\left[\rho^2\right]$, with $1/d\leq\mathcal{P}\leq1$ \cite{bar09}. This is closely related to the linearized version of the von Neumann entropy.  If $\mathcal{P}=1$ ($\mathcal{P}=1/d$), the system is in a pure (maximally mixed) state. For our purposes of contrasting RWA and non-RWA descriptions, the purity is more suitable than the full entropy since the former allowed us to get analytic and exact expressions. For the initial conditions given by Eqs.~\eqref{eq:i2qb} and~\eqref{eq:i2md}, we obtain 
\begin{eqnarray}
\mathcal{P}_{j}(t)=\text{Tr}\left[\rho_{AB_j}^2(t)\right]=\frac{1}{4}+2\Gamma_{j}(t),
\label{eq:purAB}
\end{eqnarray}\\
with 
\begin{eqnarray}
\Gamma_{j}(t)&=&\left|\rho_{AB,eeeg_j}(t)\right|^2+\left|\rho_{AB,eege_j}(t)\right|^2+\left|\rho_{AB,eegg_j}(t)\right|^2 \nonumber \\
&&+\,\left|\rho_{AB,egge_j}(t)\right|^2+\left|\rho_{AB,eggg_j}(t)\right|^2\nonumber\\ &&+\left|\rho_{AB,gegg_j}(t)\right|^2.
\label{eq:Gamma}
\end{eqnarray}
Explicitly, the function $\Gamma_{j}(t)$ reads 
\begin{widetext}
\begin{eqnarray}
\Gamma_{j}(t)=4e^{-\left[f_{j_+}(t)+f_{j_-}(t)+2\gamma t\right]}+e^{-4\left[f_{j_+}(t)+\gamma t\right]}+e^{-4\left[f_{j_-}(t)+\gamma t\right]},
\label{eq:Gamma1}
\end{eqnarray}
with
\begin{eqnarray}
f_{j_\pm}(t)=16\lambda_{j_\pm}^2\left[\cosh(2r_{j_\pm})-\sinh(2r_{j_\pm})\cos(\Omega_{j_\pm} t)\right]\sin ^2\left(\frac{\Omega_{j_\pm} t}{2}\right).
\label{eq:fj}
\end{eqnarray}
\end{widetext}

\begin{figure}[h!]
\centering
\includegraphics[scale=0.67]{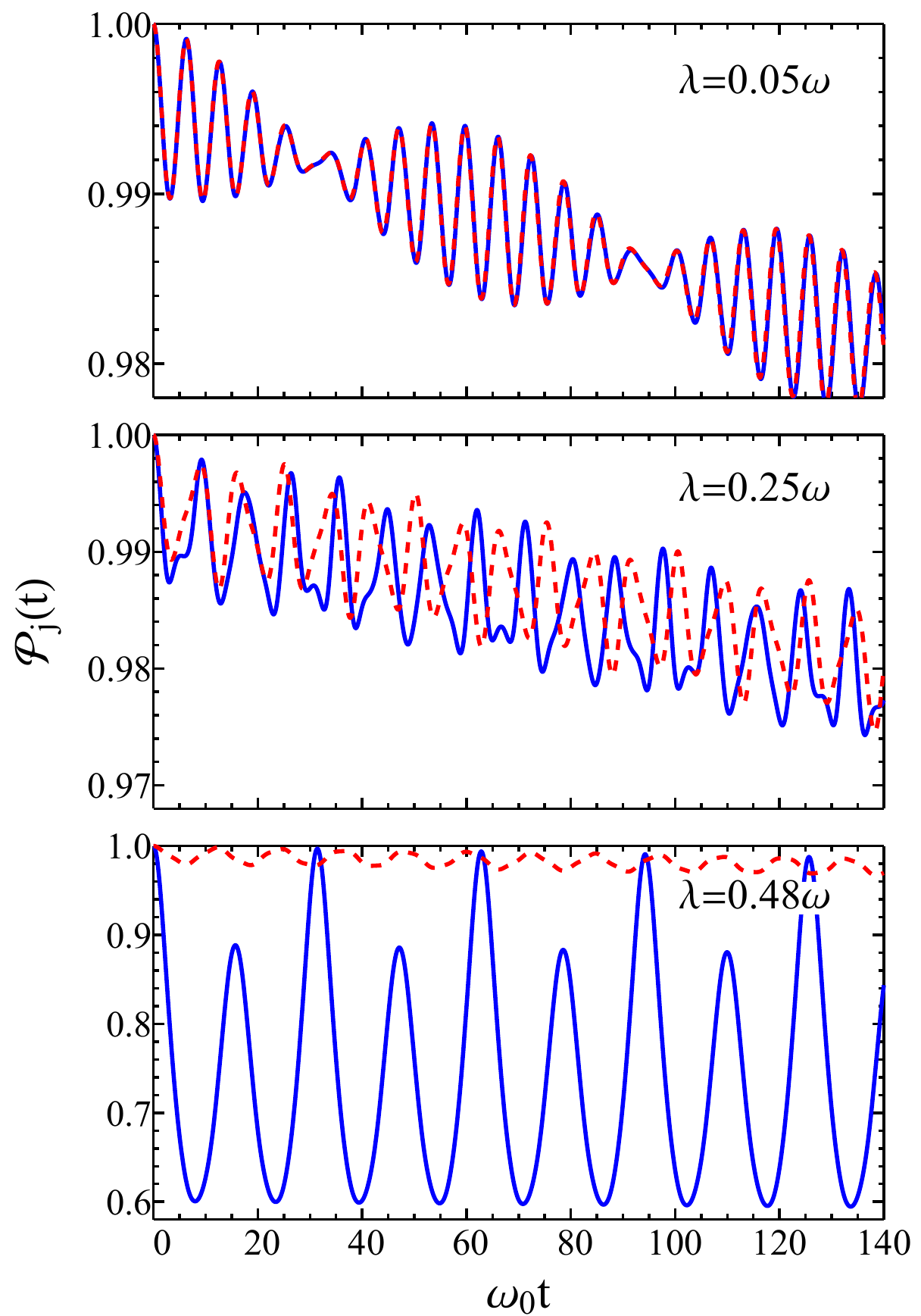}
\caption{\label{fg:purAB} (Color online) Purity $\mathcal{P}_{j}$ of the two-qubit subsystem as function of the dimensionless time $\omega_0 t$ for different values of mode coupling strengths $\lambda$. Solid blue lines represent the purity for the complete model ($j=1$), whereas dashed red lines represent the purity under the RWA model ($j=2$). 
We used $\omega_0=\omega$, $g=0.025\ \omega_0$, and $\gamma=5\times10^{-5}\ \omega_0$.}
\end{figure}

Figure~\ref{fg:purAB} compares the dynamics of $\mathcal{P}_j(t)$ for each model under different mode coupling regimes. Evidently, the purity is maximum at $t=0$ as the initial state of the qubits is $\ket{++}$  and not entangled with the modes. The first thing to be noticed is that there is a clear competition between the Markovian dynamics induced by the dephasing baths and the non-Markovian dynamics induced by the mode-mode interaction \cite{car15}. To be more precise, the oscillations appear as the result of the latter while the envelop (purity damping) is caused by the baths. Such features are caused by exponentials of multiples of $-\gamma t$ and $f_{j_\pm}(t)$ in Eq.~\eqref{eq:Gamma1}, respectively. Purities for both the complete and the RWA models turned out to be independent on the initial coherent states of the modes. This is so because $\alpha$ and $\beta$ can be eliminated from the dynamics via a time-independent unitary transformation on the modes (basis transformation).

Let us now closely examine the dependence of the two-qubit purity on mode-mode coupling strength $\lambda$. From Fig.~\ref{fg:purAB}, we can  see that, for moderate couplings ($\lambda\approx 0.25\ \omega_0$), the predictions of the RWA and non-RWA models already disagree considerably. Although they both have similar orders of magnitude, the oscillations are not in phase anymore (compared to small $\lambda$). This is a direct consequence of the deviations in $\Omega_{j_\pm}$ caused essentially by the squeezing parameter $r_{j_\pm}$. In addition, as $\lambda$ becomes larger (approaching the limit $\omega/2$), the RWA model fails miserably to predict the correct phases and amplitudes. It is noticeable that in the full non-RWA model the amplitude of purity oscillation is much larger than the RWA prediction. This can be attributed to content of the square brackets in Eq.~\eqref{eq:fj}. For $j=1$ (non-RWA) it is an oscillating function while for  $j=2$ it is constant and equals one. 
\subsection{\label{subsec:ent}Entanglement of the qubits}
We now turn our analysis to the two-qubit entanglement generation in the studied setup. Some quantum information tasks, such as quantum teleportation \cite{ben93}, need the handling of large amounts of entanglement to be performed properly, so that it is essential to determine how entangled a certain system is. In this work, we use the concept of entanglement of formation of an arbitrary two-qubit mixed state \cite{woo98,horo09}. First, one defines the so-called concurrence function $\mathcal{C}(\rho) =\text{max}\{0, \sqrt{\epsilon_1}-\sqrt{\epsilon_2}-\sqrt{\epsilon_3}-\sqrt{\epsilon_4}\}$, where $\epsilon_i$s are the eigenvalues in decreasing order of $\rho\sigma_{y_A}\sigma_{y_B}\rho^*\sigma_{y_A}\sigma_{y_B}$, $\rho^*$ is the complex conjugate of $\rho$, and $\sigma_y$ the $y$-Pauli matrix. Then, the entanglement of formation of $\rho$ can be defined as
\begin{eqnarray}
 \mathcal{E}_F(\rho)=h\left(\frac{1+\sqrt{1-\mathcal{C}(\rho)^2}}{2}\right),
 \label{eq:eof}
\end{eqnarray}
where $h(x)=-x\log_2x-(1-x)\log_2(1-x)$. Both concurrence and entanglement of formation are equal to zero (unity) for a separable (maximally entangled) state. 

\begin{figure*}
\includegraphics[width=\textwidth]{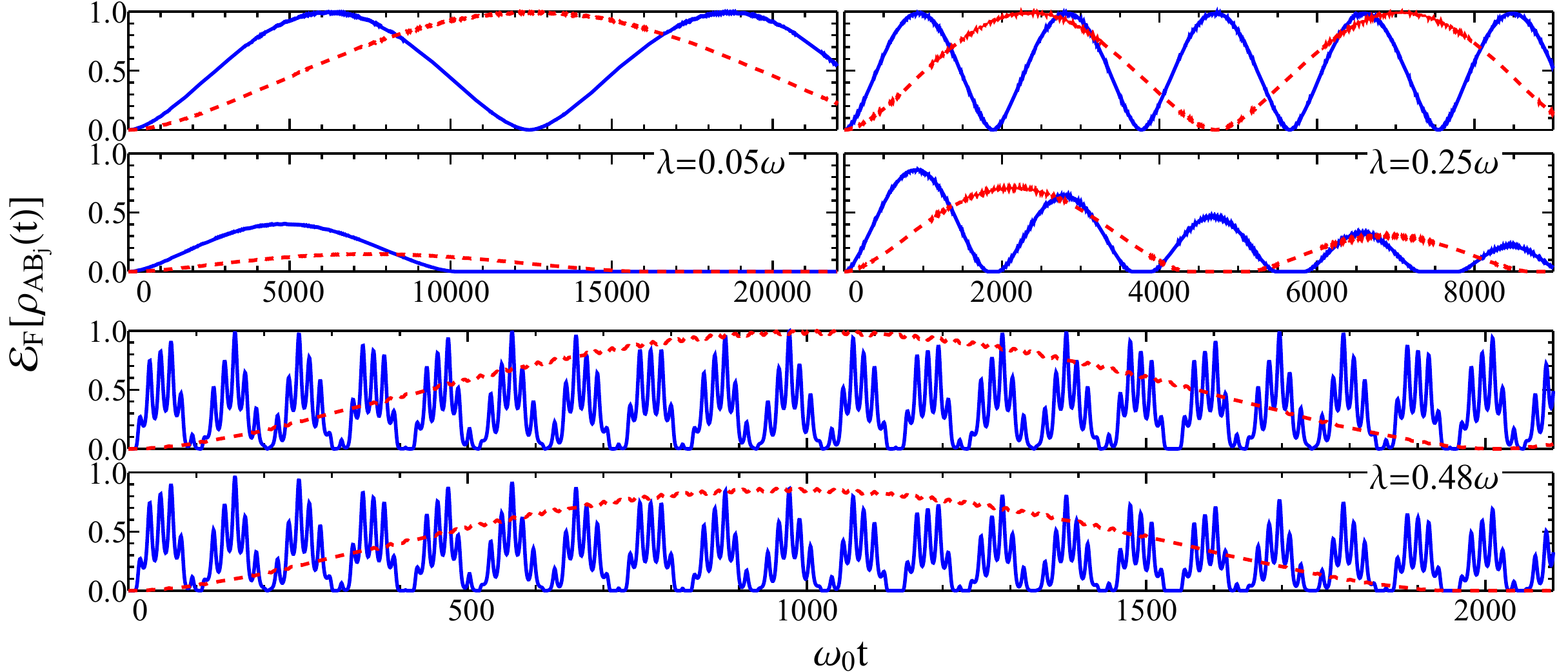}
\caption{\label{fg:entAB} (Color online) Long-time behavior of the entanglement of formation $\mathcal{E}_F$ of the two-qubit subsystem as function of the dimensionless time $\omega_0 t$ for different values of mode coupling strengths $\lambda$ and qubit dephasing rates $\gamma$. Solid blue lines represent $\mathcal{E}_F$ for the complete model ($j=1$), whereas dashed red lines represent $\mathcal{E}_F$ under the RWA ($j=2$).  
\textbf{First and second rows}: $\lambda=0.05\ \omega$ (left) and $\lambda=0.25\ \omega$ (right) with $\gamma=0$ (first row) and $\gamma=5\times 10^{-5}\ \omega_0$ (second row). \textbf{Third and forth rows}: $\lambda=0.48\ \omega$ with $\gamma=0$ (third row) and $\gamma=5\times 10^{-5}\ \omega_0$ (forth row). The remaining parameters are $\omega_0=\omega $, $g=0.025\ \omega_0$, and $\alpha=\beta=2$.}
\end{figure*} 
\begin{figure}[h!]
\centering
\includegraphics[scale=0.67]{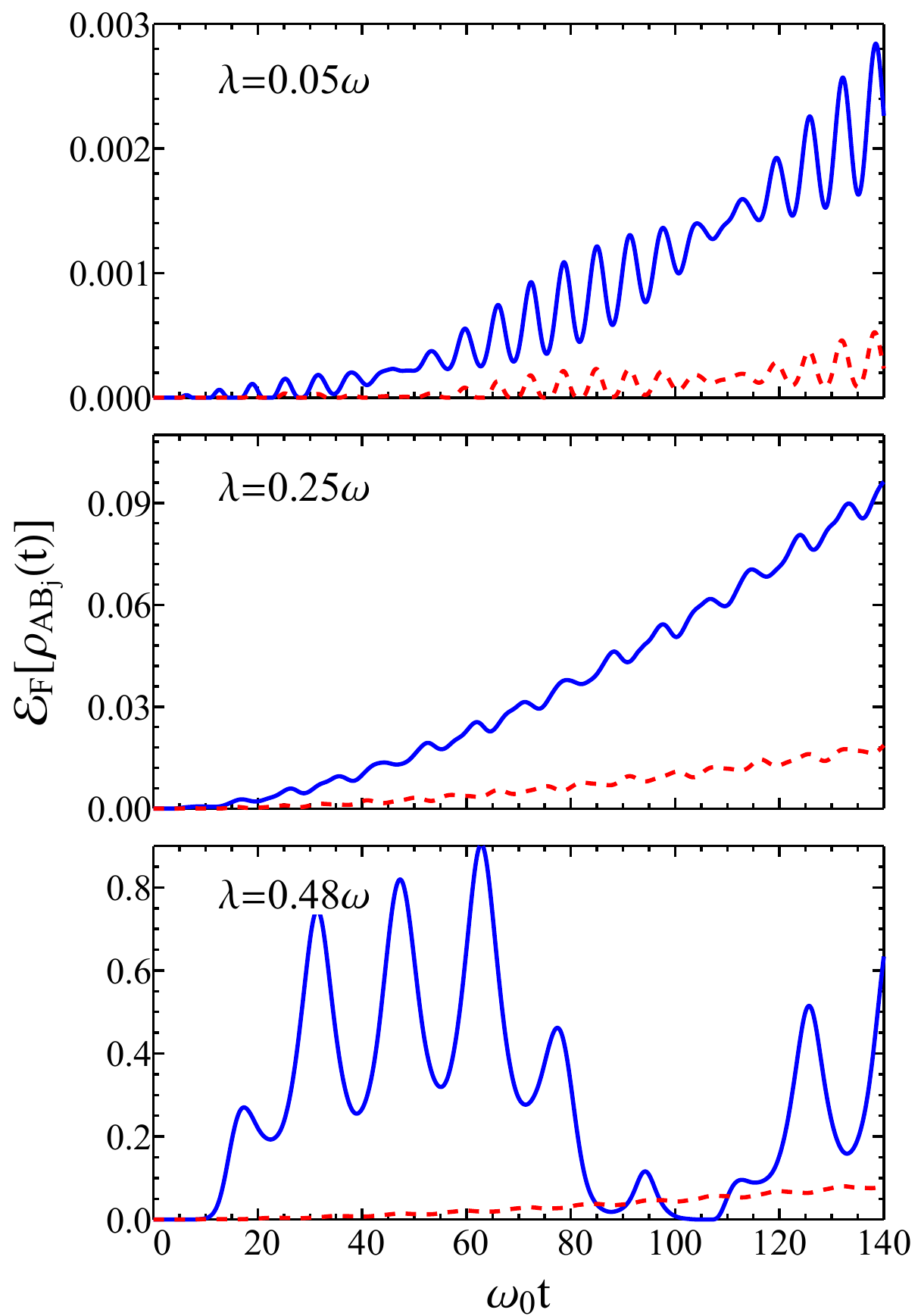}
\caption{\label{fg:entABii} (Color online) Short-time behavior of the entanglement of formation $\mathcal{E}_F$ of the two-qubit subsystem as function of the dimensionless time $\omega_0 t$ for different values of mode coupling strengths $\lambda$. Solid blue lines represent $\mathcal{E}_F$ for the complete model ($j=1$), whereas dashed red lines represent $\mathcal{E}_F$ under the RWA ($j=2$). 
The remaining parameters are $\omega_0=\omega$, $g=0.025\ \omega_0$, $\gamma=5\times10^{-5}\ \omega_0$ and $\alpha=\beta=2$.}
\end{figure}

Figure~\ref{fg:entAB} compares the dynamics of $\mathcal{E}_F[\rho_{AB_j}(t)]$ for different values of qubit dephasing rates $\gamma$ and mode coupling strengths $\lambda$. As expected, $\mathcal{E}_F$ starts from zero as the initial state is separable. Then $\mathcal{E}_F$ oscillates and reaches its maximum values for the conditions taken in the first and third rows ($\gamma=0$). In this scenario, the state of the two-qubit system periodically changes from separable to maximally entangled states in the absence of dephasing. In the second and forth rows, on the other hand, such oscillations are damped due to the non-null dephasing rate ($\gamma=5\times10^{-5}\ \omega_0$), and the capability of producing higher peaks of $\mathcal{E}_F$ is enhanced in the strong coupling regime. Thus, the non-Markovian aspect of the dynamics acts as an entanglement generator, whereas the Markovian part tends to destroy it over long times. 

Another feature present in Fig.~\ref{fg:entAB} is that the main frequency of oscillation of $\mathcal{E}_F$ increases with $\lambda$, independently of the model. However, for the full model, the generation of entanglement is evidently faster than in the RWA. This might be interesting for a scenario where entanglement needs to be preserved in the presence of strong dephasing. Fast generation of entanglement has recently attracted interest of the community and has already been proposed in other setups \cite{peng12,bel13,qiu13}. This can be obtained, for example, if two non-interacting qubits are weakly coupled to a common Ohmic bath \cite{marz09,marz10}.

By examining further the short-time behavior of $\mathcal{E}_F$ (Fig.~\ref{fg:entABii}), one can see that it essentially oscillates with the same fast frequencies of the two-qubit purity (Fig.~\ref{fg:purAB}). This is indeed expected since both quantities are indirectly related to the local entropies for each qubit subsystem. Differently from what is observed in Fig.~\ref{fg:purAB}, the deviations in $\mathcal{E}_F$ for both models become evident even for small values of mode coupling strengths (e.g. $\lambda=0.05\ \omega$). This is a consequence of the complexity of $\mathcal{E}_F$ compared to $\mathcal{P}_j$. To evaluate the former it is necessary density matrix diagonalization and application of logarithmic functions while for the latter it is just necessary to square it and trace. Also, since what justifies the RWA is precisely first order perturbation theory \cite{boiuna}, valid for short times, we indeed expect that the stronger the $\lambda$ the shorter the time range for which RWA provides a satisfactory answer, and this is clearly seen from the plots in Figs.~\ref{fg:entAB} and~\ref{fg:entABii}. Therefore, one can conclude that the dynamics of entanglement is more sensible to the variations of $\lambda$ than purity is.
\section{\label{sec:conc}Conclusion}
We have provided an illustrative example where the inadequacy of the RWA can be analytically investigated. The system is composed of quantum two-level systems and harmonic oscillators, both ubiquitous in controlled quantum systems such as trapped ions or circuit QED. In particular, we have focused on a model comprised of two qubit-mode subsystems which are brought into interaction through their harmonic coordinates. We have performed exact diagonalization of the full model and its RWA version, and analytically solved the master equation for initial states of interest. 

We then have shown that the modes coupling strength $\lambda$ plays a fundamental role on the variety of responses of the qubits as displayed by state purity and entanglement. The predictions of the complete model and the RWA model for short times and $\lambda \approx 0.05\ \omega$ agree well for purity but not for entanglement dynamics. Also, even for short times, as soon as $\lambda \approx 0.25\ \omega$, purity is no longer described correctly by the RWA. At longer times, when two-qubit entanglement is fully generated, the failure of the RWA becomes more noticeable as it considerably reduces the main frequency of entanglement oscillations. Our study also showed that, the stronger the modes are coupled, the larger the reduction of the degree of purity at short time scale and the faster generation of entanglement. Moreover, we verified the competition between Markovian dephasing on the qubits and non-Markovianity induced by the modes coupling constant in conformity with Ref. \cite{car15}. 
\begin{acknowledgements}
P.C.C. would like to thank FAPESP for the support through Grant No. 2012/12702-7. W.S.T would like to thank CAPES for the current scholarship. F.L.S. acknowledges participation as a member of the Brazilian National Institute of Science and Technology of Quantum Information (INCT/IQ). F.L.S. also acknowledges partial support from CNPq under Grant No. 307774/2014-7.
\end{acknowledgements}
\appendix*
\section{}
The time-dependent factors $\gamma_{mn_j}(t)$ that appear in the qubits density matrix elements $\rho_{AB,mn_j}(t)$  in Eq.~\eqref{eq:kmn} are explicitly 
\begin{eqnarray}
\gamma_{eeeg_j}(t)&=&e^{-\left[i\left(\omega_{0,j}+\chi_j\right)+\gamma\right]t}e^{2i\text{Im}\left[A_{j_+}(t)+B_{j_-}(t)\right]}, \nonumber \\
\gamma_{eege_j}(t)&=&e^{-\left[i\left(\omega_{0,j}+\chi_j\right)+\gamma\right]t}e^{2i\text{Im}\left[A_{j_+}(t)-B_{j_+}(t)\right]}, \nonumber \\
\gamma_{eegg_j}(t)&=&e^{-2\left(i\omega_{0,j}+\gamma\right)t}e^{4i\text{Im}\left[\lambda_{j_+}Z_{j}^{*}\left(1-e^{i\Omega_{j_+}t}\right)\right]}, \nonumber \\ 
\gamma_{egge_j}(t)&=&e^{-2\gamma t}e^{-4i\text{Im}\left[\lambda_{j_-}W_{j}^{*}\left(1-e^{i\Omega_{j_-}t}\right)\right]}, \nonumber \\
\gamma_{eggg_j}(t)&=&e^{-\left[i\left(\omega_{0,j}-\chi_j\right)+\gamma\right]t}e^{2i\text{Im}\left[A_{j_-}(t)-B_{j_-}(t)\right]}, \nonumber \\
\gamma_{gegg_j}(t)&=&e^{-\left[i\left(\omega_{0,j}-\chi_j\right)+\gamma\right]t}e^{2i\text{Im}\left[A_{j_-}(t)+B_{j_+}(t)\right]}, 
\label{eq:gammaj}
\end{eqnarray}
where $A_{j_\pm}(t)=\lambda_{j_+}Z^{*}_{j}-\lambda_{j_+}(Z^{*}_{j}\pm2\lambda_{j_+})e^{i\Omega_{j_+}t}$ and $B_{j_\pm}(t)=\lambda_{j_-}W^{*}_{j}-\lambda_{j_-}(W^{*}_{j}\pm2\lambda_{j_-})e^{i\Omega_{j_-}t}$ are complex functions. Information about the complex amplitudes of the initial coherent state of the modes is encoded in  $Z_j=z_{j}\cosh(r_{j_+})+z^{*}_{j}\sinh(r_{j_+})$ and $W_j=w\cosh(r_{j_-})+w^{*}\sinh(r_{j_-})$, with $z_j=\left(\alpha+\beta+2\delta_j\right)/\sqrt{2}$ and $w=\left(\beta-\alpha\right)/\sqrt{2}$. Moreover,  $\gamma_{mm_j}=1$  and, given the Hermiticity of the density matrix, the remaining phases are complex conjugates of the ones in Eq.~\eqref{eq:gammaj}. It is interesting to notice that the amplitudes of the coherent states appear in the density matrix but, as explained before, do not show up in quantities that are independent of local time-independent transformations.

From the partial trace over the bosonic modes in Eq.~\eqref{eq:rhoAB} using the total initial state $\ket{\psi(0)}=\ket{++}\ket{\alpha}\ket{\beta}$, one finds the terms \begin{eqnarray}
\!\!\!\!\!\!\!\Theta_{mn,j_\pm}(t)=\braket{Y_{m,j_\pm}(t),\xi_{j_\pm}(t)}{Y_{n,j_\pm}(t),\xi_{j_\pm}(t)},
\label{eq:scpd}
\end{eqnarray}
which are scalar products of squeezed coherent states with squeezing parameter $\xi_{j_\pm}=-r_{j_\pm}e^{-2i\Omega_{j_\pm}t}$ and amplitudes 
\begin{eqnarray}
Y_{ee,j_+}(t)&=&\left(Z_{j}+2\lambda_{j_+}\right)e^{-i\Omega_{j_+}t}-2\lambda_{j_+}, \nonumber \\
Y_{eg,j_+}(t)&=&Z_{j}e^{-i\Omega_{j_+}t}=Y_{ge,j_+}(t), \nonumber \\
Y_{gg,j_+}(t)&=&\left(Z_{j}-2\lambda_{j_+}\right)e^{-i\Omega_{j_+}t}+2\lambda_{j_+}, \nonumber \\
Y_{ee,j_-}(t)&=&W_{j}e^{-i\Omega_{j_-}t}=Y_{gg,j_-}(t), \nonumber \\
Y_{eg,j_-}(t)&=&\left(W_{j}-2\lambda_{j_-}\right)e^{-i\Omega_{j_-}t}+2\lambda_{j_-}, \nonumber \\
Y_{ge,j_-}(t)&=&\left(W_{j}+2\lambda_{j_-}\right)e^{-i\Omega_{j_-}t}-2\lambda_{j_-}.
\label{eq:Ynj}
\end{eqnarray}
Notice that in the RWA, i.e., $j=2$,  $\Theta_{mn,2_\pm}(t)$ reduces to the overlap of coherent states $\braket{Y_{m,2_\pm}(t)}{Y_{n,2_\pm}(t)}$.

\end{document}